\documentclass[showkeys,reprint,nofootinbib]{revtex4-1}
\usepackage{graphicx}
\usepackage{times}
\usepackage{mathrsfs}
\usepackage{amsmath,amssymb}

\allowdisplaybreaks
\newcommand{\eqhyp}[1]{#1\nobreak\discretionary{}{\hbox{\ensuremath{#1}}}{}}

\begin{document}
\abovedisplayskip=3pt plus 3.0pt minus 3.0pt
\abovedisplayshortskip=0.0pt plus 3.0pt
\belowdisplayskip=3pt plus 3.0pt minus 3.0pt
\belowdisplayshortskip=3pt plus 3pt minus 3pt

\title{\LARGE Electron spin resonance measurements of a demagnetizing field on~the surface of metal samples}
\author{N.A. Poklonski}
\email{poklonski@bsu.by}
\author{T.M. Lapchuk}
\author{N.I. Gorbachuk}
\affiliation{Belarusian State University, pr. Nezavisimosti 4, 220030 Minsk, Belarus}


\date{\today}

\begin{abstract}
By comparing the signals of electron spin resonance (ESR) from two crystals of a diamond (spin-labels) the demagnetizing field of the Co, Fe, and Ni samples in the shape of strongly elongated ellipsoids of revolution (disks) has been measured. The magnetic permeabilities of the metals in the external magnetic field corresponding to the ESR of the broken chemical bonds in a natural diamond irradiated with fast reactor neutrons have been determined.
\end{abstract}


\maketitle 

In ESR spectroscopy, spin-labels are used for investigation of electrical and magnetic properties of materials. For example, in [1] external spin-labels were applied to measure the electrical conductivity of nonmagnetic metals and semiconductors. A plate of ruby (Al$_2$O$_3$:Cr$^{3+}$) was coated with various electric conductors and placed on a quartz holder in the center of a rectangular $H_{102}$-resonator. In this case, the ESR signal of the Cr$^{3+}$ ions was distorted and assumed the shape of the Dyson line. The ratio of the amplitudes of the low- and high-field wings of the Cr$^{3+}$ absorption line gave the impedance of the multilayer structure in the microwave field.

In [2], nickel-foil inserts placed within a microwave resonator were used to create a gradient of the polarizing magnetic field (ESR tomography method) to investigate the distribution of paramagnetic centers inside dielectric samples.

The dynamic magnetic susceptibility of antiferromagnetic CuO-based semiconductors was studied in [3] by the method of the ESR of internal spin-labels (Cu$^{2+}$ ions). Rotation of the polarization plane of microwave oscillations by the semiconductor sample in a cylindrical $H_{111}$-resonator with scanning of the polarizing magnetic field was recorded.

As far as we know, the external spin-labels have not been used for measurement of the magnetic properties of materials. The purpose of our work is to determine the demagnetizing field\footnote{The term ``demagnetizing field'' [4] means that the magnetic field in a sample of finite dimension depends on its shape and is smaller than that in a closed toroid made of the same material. The smaller the demagnetizing factor, the stronger the magnetization of the samples.} and the magnetic permeability of metal samples by the ESR method with the aid of two spin-labels.

We recorded the ESR spectra on a RadioPAN SE/X-2543 spectrometer in the \emph{X}-band with automatic adjustment of the frequency of the microwave generator (klystron) to the frequency of the $H_{102}$-resonator. The polarizing magnetic field was modulated with frequencies of 100~kHz and 80~Hz; the modulation amplitude was 0.05~mT. An NMR magnetometer and a Hall probe were used to measure the magnitude of the polarizing field, and a frequency meter was used to measure the resonator frequency. The ESR signal of the ruby, positioned on a wall of the resonator, was used for monitoring the \emph{Q}-factor of the resonator, tuning a phase of the magnetic field modulation, and calibration of the $H_1$-component of the microwave field. The ruby in the resonator was placed so that its ESR signal could not be saturated in the range of microwave powers used (up to 50~mW). The recommended volume of paramagnetic dielectric samples for such resonators is 128~mm$^3$. The measurements were carried out at room temperature.

We investigated chemically pure polycrystalline samples of Co, Fe, Ni, W, Mo, and Ta without residual magnetization. Every sample was machined to a shape close to a strongly elongated ellipsoid of revolution (hereinafter a disk) having a diameter of $d ={}$3~mm and a thickness of $t ={}$0.1~mm; the characteristic dimensions of the surface irregularities did not exceed 10~$\mu$m. The choice of the dimensions of the metal samples was dictated by the necessity of minimizing the absorption of the electrical component of the microwave field by them. The metal disks were oriented in the resonator so that their plane was parallel to the directions of both the external field induction $B$ and the magnetic component of the microwave field $H_1$. With this orientation, the demagnetization factor of the disk is minimal, and it can be calculated (see below).

For measurement of the demagnetizing fields, point spin-labels on the surface of ferromagnetic (Co, Fe, Ni) and nonmagnetic (W, Mo, Ta) metals were used. As the spin-labels we took two 0.4 and 0.2~mm-diameter natural diamonds of round shape irradiated by a fluence of 3$\cdot$10$^{20}$ fast reactor neutrons$/$cm$^2$. The paramagnetism of the irradiated diamond is due to broken chemical bonds. The isotropic \emph{g}-\hspace{0pt}factor was determined in the usual way from the measured frequency $\nu$ of the microwave resonator and the magnetic induction $B$ of the external polarizing field [5].

To fix the samples, a cylindrical quartz rod was used in which a cut was made to provide the positioning of the metal disk and two spin-labels along the vertical axis of the $H_{102}$-resonator in the antinode of the magnetic component of the microwave field.

The larger spin-label was pasted onto the surface of the metal disk at its center and the smaller one on the quartz holder at a distance of 4.5~mm from the disk edge. The necessity of having a larger number of paramagnetic centers on the diamond pasted onto the metal disk is attributable to the inhomogeneity of the demagnetizing field leading to a decrease in the amplitude of the ESR signal. The smaller spin-label was pasted onto the quartz holder at a distance of 6~mm from the disk center to minimize the influence of the demagnetizing field of the disk on it and not to lie outside the boundaries of the effective center of the resonator.

The sequentially recorded signals of the absorption of the microwave field by the metal disks that were alternately placed (on the quartz holder) at the center of the $H_{102}$-\hspace{0pt}resonator are presented in Fig.~1. It is seen that the ferromagnetic resonance [6, 7] with characteristic wide lines is observed for Co, Fe, and Ni and is not manifested for W, Mo, and Ta. Note that the eigenfrequency and the \emph{Q}-factor of the resonator change when the ferromagnetic disk is placed into the antinode of the magnetic component of the microwave field.

\begin{figure}
\noindent\hfil\includegraphics{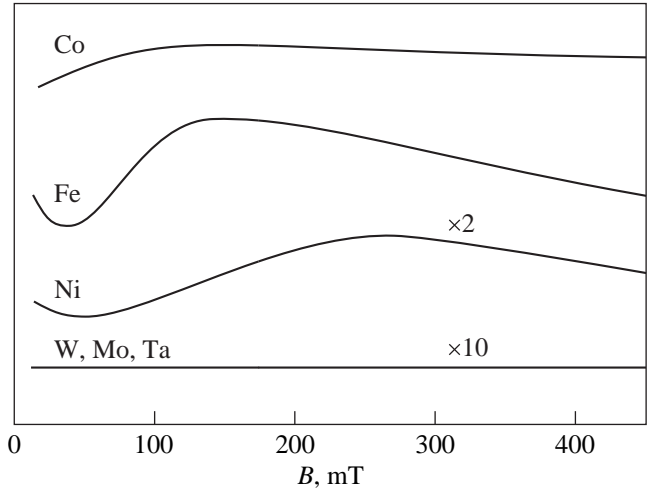}
\caption{Signal of absorption of the microwave field of metal samples
(designated by the chemical symbols of the elements) of disk shape in
the magnetic field. The numerals are amplifications of the signal in recording.
The polarizing field $B$ and the $H_1$-component of the microwave
field are in the disk plane.}\label{fig:01}
\end{figure}

Figure 2 shows the ESR signal of the diamond spin-labels which were spaced 6~mm apart on the quartz holder without metal samples (Dia$^\text{\dag}$ signal) and the ESR signals from the same spin-labels, but in the presence of the corresponding metal disks. It is evident that without the ferromagnetic metal in the resonator the ESR spectrum of two diamonds represents a single line of width 0.2~ mT, which is sum of their signals. The measured values $B = \mu_0H ={}$332.1~mT, where $\mu_0$ is the magnetic constant, and $\nu ={}$9.3079~GHz gave $g ={}$2.0026${}\pm{}$0.0002. Under the action of the demagnetizing field $H_\text{d}$ of the disk from Co, Fe, and Ni, the combined line of the ESR of the diamonds is split into two lines. The narrow line ($\sim{}$0.3~mT) corresponds to the diamond located at a distance of 6~mm from the center of the disk, and the broadened one is attributed to the diamond located directly on the surface of the disk, above its center. The influence of the demagnetizing field of the disk on the diamond located on the magnetized disk is seen as displacement of its ESR line by $\delta B = \mu_0H_\text{d}$ to the side of higher magnetic fields. The resonant values of the magnetic induction $\mu_0H(\text{Co}) \eqhyp\approx{}$332.9~mT, $\mu_0H(\text{Fe}) \eqhyp\approx{}$332.7~mT, and $\mu_0H(\text{Ni}) \eqhyp\approx{}$332.4~mT correspond to the center of the ESR line of the diamond which is out of the field of the ferromagnetic disk. The shift of the ESR line of the diamond on the disk to the region of higher magnetic fields with respect to that of the diamond on the quartz holder comes to $\delta B(\text{Co}) \eqhyp\approx{}$18.4~mT, $\delta B(\text{Fe}) \eqhyp\approx{}$16.3~mT, and $\delta B(\text{Ni}) \eqhyp\approx{}$3.4~mT. For the W, Mo, and Ta disks, splitting of the combined line from two diamond samples is not observed. A transition from the polarizing magnetic field modulation frequency of 100~kHz to that of 80~Hz did not influence the shift $\delta B$ of the spin-label absorption line.

\begin{figure}
\noindent\hfil\includegraphics{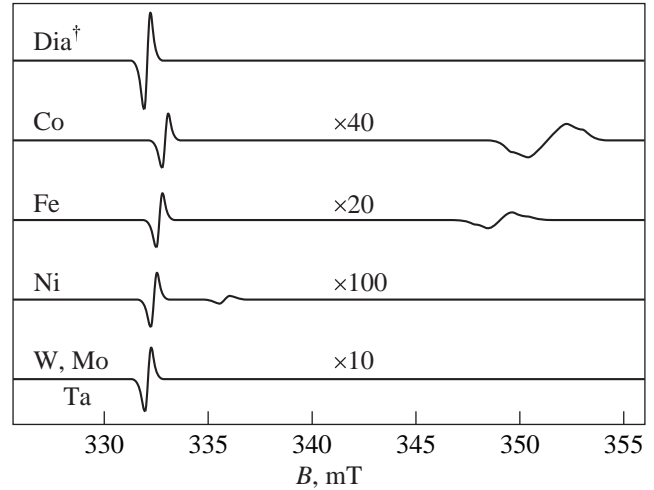}
\caption{ESR spectra of two diamond spin-labels. The Dia$^\text{\dag}$ spectrum belongs
to diamonds that are 6~mm apart on the quartz holder in the center
of the $H_{102}$-resonator. The spectra denoted by the chemical symbols of
the elements belong to the spin-labels, one of which is located on the
surface of the metal disk above its center.}\label{fig:02}
\end{figure}

According to [4, 6--9], the demagnetizing field in the oblate ellipsoid of revolution made of an isotropic metal is the following:
\begin{equation}\label{eq:01}
   H_\text{d} = NI = N\chi H_\text{i} = N(\mu_\text{r} - 1)H_\text{i},
\end{equation}
where $N$ is the demagnetization factor, $I$ is the magnetization, $\chi$ is the magnetic susceptibility, $H_\text{i} = H - H_\text{d}$ is the magnetic field intensity in the metal, and $\mu_\text{r}(H) = 1 + \chi(H)$ is the metal relative permeability depending on the external magnetic field $H$.

For the oblate ellipsoid of revolution of diameter $d$ and thickness $t = kd$, where $k < 1$, with the external magnetic field being parallel to the sectional plane of maximum area, the demagnetization factor is [8]
\begin{equation}\label{eq:02}
   N = \frac{k}{2(1 - k^2)^{3/2}}\biggl[\frac{\pi}{2} - k(1 - k^2)^{1/2} - \arcsin k\biggr].
\end{equation}

When $t \ll d$ (for the disk), Eq. (2) yields $N \approx \pi t/4d$. For the investigated disk-shaped metal samples $k = t/d \approx{}$0.033 we have $N \approx{}$0.025.

While eddy currents excited in the disk by the modulation field during ESR measurements can be neglected, the tangential component of the magnetic field intensity is continuous on the boundary of the ferromagnetic disk and the diamond. Then the intensity of the demagnetizing field on the disk surface $H_\text{d} = H - H_\text{i}$, and with allowance for Eq. (1) we obtain
\begin{equation}\label{eq:03}
   H_\text{d} = \frac{N(\mu_\text{r} - 1)}{1 + N(\mu_\text{r} - 1)} H.
\end{equation}

Thus, the distance between the centers of the ESR lines of two diamond samples $\delta B = \mu_0H_\text{d}$, where $H_\text{d}$ is determined from Eq. (3) with allowance for Eq. (2).

We have found experimentally that $\delta B = \mu_0H_\text{d}$ and, consequently, $\mu_\text{r}(H)$ depends appreciably (especially for Fe) on the heat treatment of metals. The reason is that for the region of the ESR magnetic fields of the diamond irradiated with neutrons all the domains in the ferromagnetic disks are oriented along the external field. Moreover, the manifestation of defects of the structure of the metal samples in $\mu_\text{r}(H)$ is not ``shaded'' by the influence of the reorientation of the magnetic domains, as in the region of weak magnetic fields [6--9].

Calculations of the relative permeability of the materials of the metal disks by Eq. (3) with allowance for $\mu_0H_\text{d}$ for Co, Fe, and Ni yield $\mu_\text{r}(\text{Co}) \approx{}$3.3, $\mu_\text{r}(\text{Fe}) \approx{}$3.1, and $\mu_\text{r}(\text{Ni}) \eqhyp\approx{}$1.4 at an external magnetic field induction of $B = \mu_0H \eqhyp\approx{}$332~mT and a field modulation frequency of 100~kHz.

Note that, according to (3), the value of $N(\mu_\text{r} - 1)$ can be determined from the measured values of $H_\text{d} = \delta B/\mu_0$ and $H$. Then the ratio of the relative magnetic permeability of the sample $\mu_\text{d}$ with the demagnetization factor $N$ to the permeability of the material of the sample $\mu_\text{r}$ is determined from Arkad'ev's formula [7]:
\[
   \frac{\mu_\text{d}}{\mu_\text{r}} = 1 - \frac{H_\text{d}}{H} = \frac{1}{1 + N(\mu_\text{r} - 1)}.
\]

Thus, by the ESR method of spin-labels (two crystals of diamond) the demagnetizing field and magnetic permeability of the samples of the ferromagnetic metals Co, Fe, and Ni in the form of disks (strongly elongated ellipsoids of revolution) have been measured. The established possibility of experimental determination of $\mu_\text{r}$ for the samples of canonical form [4, 6, 8], when the demagnetization factor $N$ can be calculated, is important for both the investigation of the nature of the magnetic susceptibility of materials [10--12] and applications [13, 14].

We thank N.M.~Lapchuk and S.A.~Vyrko for discussion of the work.



%
\end{document}